\documentclass[aps,pra,tightenlines,twocolumn,superscriptaddress,floatfix]{revtex4}

\usepackage[utf8]{inputenc}
\usepackage[T1]{fontenc}
\usepackage{setspace}
\usepackage{amsmath}
\usepackage{graphicx}
\usepackage{float}
\usepackage{microtype}
\usepackage{amssymb}
\usepackage{amsthm}
\usepackage{physics}

\newtheorem{mylemma}{Lemma}
\usepackage{tikz}
\usepackage{hyperref}
\usepackage{pgf-pie}
\usepackage{graphicx}
\usepackage{pdfpages}
\usepackage{color}
\usepackage{subcaption}
\usepackage[]{algorithm2e}
\usepackage{bbold}
\usepackage{mathtools}
\usepackage{enumerate}
\usepackage{tcolorbox}
\usepackage{adjustbox}
\usepackage{floatflt}
\usepackage{comment}
\usepackage{svg}
\usepackage[OT1]{fontenc}
\newcommand{\M}{\mathcal{M}}
\newcommand{\sumkl}{\sum_{k,l = 0}^{d-1}}

\newcommand{\Pkl}{P_{k,l}}
\newcommand{\Okl}{\Omega_{k,l}}

\begin{document}

\title{Almost complete solution for the NP-hard separability problem of Bell diagonal qutrits}

\author{Christopher Popp}
\email{poppchristopher@pm.me}
\affiliation{University of Vienna, Faculty of Physics, Währingerstrasse 17, 1090 Vienna, Austria}
\author{Beatrix C. Hiesmayr}
\email{Beatrix.Hiesmayr@univie.ac.at}
\affiliation{University of Vienna, Faculty of Physics, Währingerstrasse 17, 1090 Vienna, Austria}


\begin{abstract}
  With a  probability of success of $95 \%$ we solve the separability problem for Bell diagonal qutrit states with positive partial transposition (PPT).
  The separability problem, i.e. distinguishing separable and entangled
  states, generally lacks an efficient solution due to the existence of bound
  entangled states. In contrast to free entangled states that can be used for
  entanglement distillation via local operations and classical communication,
  these states cannot be detected by the Peres-Horodecki criterion or PPT
  criterion. We analyze a large family of bipartite qutrit states that can be separable, free entangled or bound entangled. Leveraging a
  geometrical representation of these states in Euclidean space, novel methods are presented
  that allow the classification of separable and bound entangled Bell diagonal
  states in an efficient way. Moreover, the classification allows the precise determination of relative
  volumes of the classes of separable, free and bound entangled states. In detail, out of all Bell diagonal PPT states $81.0 \%\pm0.1\%$ are
  determined to be separable while $13.9\pm0.1\%$ are bound entangled and only $5.1\pm0.1\%$ remain unclassified. Moreover, our applied criteria are compared for their  effectiveness and relation as detectors of bound entanglement, which reveals that not a single criterion is capable to detect all bound entangled states.
\end{abstract}

 \flushbottom
\maketitle

\section{Introduction}
Quantum information theory can be understood as the study of processing tasks
that can be accomplished using quantum mechanical systems
\cite{nielsen}. Quantum technology uses quantum phenomena for improved or new
technological applications. Prominent fields include quantum computing, quantum
communication, quantum simulation, quantum metrology or quantum
cryptography. Most of existing experimental and theoretic applications of
quantum technology are based on two-level quantum systems, called
``qubits''($d=2$). Recently however, interest in higher dimensional systems,
i.e. ``qutrits'' ($d=3$) or ``qudits'' (general $d$) is growing
\cite{highDimQCReview, qditsQCWang}. Although experimentally more challenging,
qudits are expected to have certain advantages and show new phenomena. This work
wants to explore one of these phenomena.
\\
Entanglement is one of the main resources to realize information processing with
better performance than any classical system and has been shown to
exist in many physical systems and is intensively explored for potential
applications ranging from quantum teleportation to the detection of cancer in
human beings \cite{Moskal_2016, jpet, 3photonDec, comptonMUB}. Despite its
theoretical and practical relevance, there is no general method to determine
whether a bipartite quantum state is entangled or not, in which case the state is
separable. This problem is strongly connected to a special form of entanglement
called ``bound entanglement'' for which a first example has been found by the
Horodecki family in 1998 \cite{distillation} and has been studied since
then, e.g. Refs.~\cite{Halder_BE, lockhart_e, Bruss_BE, Slater2019JaggedIO,choi, Chru_EW, BaeLinkingED,
  Bae_structPhysApprox, Korbicz2008StructuralAT, Huber_hdent,
  augus_optEw}, and has been experimentally verified with photons entangled in the orbital momentum degrees of freedom~\cite{hiesmayrLoeffler}. Often, several
weakly entangled states can be transformed via local operations and classical
communication (LOCC) to a smaller set of strongly entangled states. This process
is called entanglement distillation \cite{distillation} and is of high relevance
for practical applications that require strong distillable or ``free''
entanglement as a resource. For bound entangled states, this distillation is not
possible. However, they can be generated by mixing several free entangled states,
binding or preventing their entanglement to be further used in applications that
require distillation, thus motivating the name. In order to avoid operations
that would bind the resource of entanglement in bound entangled states through
this irreversible mixing process, it is important to have information
about the structure of bound entangled states.  While free entanglement can be
efficiently detected by the the Peres-Horodecki criterion (also known as PPT
criterion) \cite{peres,horodeckiCriterion} in a bipartite system, the detection
of bound entanglement has been proved to be NP-hard for bipartite systems and
thus intractable for higher dimensions \cite{nphard}. Up to now, there is no
complete understanding about why this special form of entanglement exists, what
its existence implies for nature or what its exact capabilities with respect to
quantum information theoretic tasks are. Moreover, there is no general method
known to construct bound entangled states.
\\
In this paper we investigate mixtures of maximally entangled, bipartite,
$3\times3$-dimensional (``Bell'')-states. One significant subset of this class of
Bell diagonal states has been shown \cite{baumgartner1,
  hiesmayrLoeffler,baumgartnerHiesmayr,baumgartnerHiesmayr2} to be especially
useful to analyze (bound) entanglement due to a representation, which allows a
precise geometrical interpretation of the system. Depending on the mixing
probabilities of pure maximally entangled states, the mixed state can be
separable, bound entangled or free entangled. Although geometric
properties allow analytical insight in the structure of entanglement
\cite{baumgartner1}, the problem of entanglement classification, especially the
detection of bound entanglement, remains a hard problem. Several methods to
detect bound entangled states exist, \cite{choi, Chru_EW, BaeLinkingED,
  Bae_structPhysApprox, Korbicz2008StructuralAT, Huber_hdent,
  augus_optEw} but these methods based on specifically constructed
non-decomposable entanglement witnesses or other constructions are strongly
limited in the number of states they can detect.  K.~Zyczkowski numerically
analyzed \cite{volOfSep1,volOfSep2} general higher dimensional systems with
respect to the separability problem and entanglement classification and
determined approximate relative volumes of separable and bound entangled states
for different dimensions. Among other results it was shown that the relative
volume of separable and bound entangled states decreases exponentially with the
dimension of the system. Recently, B. C.~Hiesmayr analyzed mixtures of Bell
states for $d=3$ to determine relative volumes of separable and (bound)
entangled states and to detect relevant substructures in
the set of bound entangled states \cite{hiesmayr1}.\\
This work intends to set up a numeric framework that allows the efficient and
precise determination of the entanglement class of Bell diagonal states in
general dimension and to apply it to the case $d=3$. An alternative, more
precise and efficient way to generate and analyze states in order to estimate
the relative volumes of entanglement classes is presented. New methods are
developed to efficiently solve the separability problem in the considered system
for almost all states. Finally, the methods are compared in their effectiveness
for the detection of bound entanglement.\\
The paper is organized as follows: First, we define the systems to be analyzed
and several subsets of relevance for the investigation. Then, a method to
efficiently generate states of that system based on random sampling is presented
and shown to have significant advantages compared to the methods used in
previous related analyses \cite{hiesmayr1}. We show how symmetries of the system
can be numerically generated and leveraged for entanglement classification. We
then give a short overview over implemented and used methods for the detection
of (bound) entangled and separable states, including a new numerical sufficient
criterion for separability. Finally, these methods are applied to the system for
$d=3$ and some relevant subset, allowing to efficiently determine the
entanglement class of $95\%$ of all PPT states.

 \section{Methods}
 In this paper we analyze mixtures of maximally entangled orthonormal Bell
 states $\ket{\Okl} \in \mathcal{H} = \mathcal{H}_1 \otimes \mathcal{H}_2$ for
 $k, l = 0, 1, \cdots , (d-1)$ in the Hilbert space $\mathcal{H}$ of a bipartite
 system containing two qudits of dimension $d$. Suppose Alice and Bob share a
 bipartite maximally entangled state
 $ \ket{\Omega_{00}} \equiv \frac{1}{\sqrt{d}} \sum_{i = 0}^{d-1} \ket{ii}$. A
 basis of orthonormal Bell states $\ket{\Okl}$ can be generated by applying the
 Weyl operators \cite{teleportingWeyl}
 $ W_{k,l} \equiv \sum_{j=0}^{d-1}w^{j \cdot k} \ket{j} \bra{j+l \pmod d},~w =
 e^{\frac{2 \pi i}{d}}, $ to one of the subsystems, without loss of generality
 Alice's qudit:
 $\ket{\Okl} \equiv W_{k,l} \otimes \mathbb{1}_d \ket{\Omega_{00}}$. The density
 matrices of these pure basis states are then the ``Bell projectors''
 $P_{kl} \equiv \ket{\Okl}\bra{\Okl}$. Mixtures of these states define the
 system of interest $\M_d$ in this work:
\begin{gather}
\M_d \equiv \lbrace \rho = \sumkl c_{k,l} P_{k,l}~ |~
\sumkl c_{k,l} = 1, c_{k,l} \geq 0  \rbrace
\end{gather}
$\M_d$ has also been named ``magic simplex''
\cite{baumgartner1,baumgartnerHiesmayr,baumgartnerHiesmayr2}, referring to the
``magic Bell basis'' introduced by Wootters and Hill \cite{magicBellBasis} and
the fact that it can be represented as simplex in real space, identifying the
mixing probabilities $c_{k,l}$ as coordinates with the Bell projectors $P_{k,l}$
lying at the vertices of the simplex. A special property of this set is that the
reduced states, i.e. the partial trace with respect to one of the subsystems, of
all states in $\M_d$ are maximally mixed. They are therefore said to be
``locally maximally mixed'' states. Note that for $d>2$, $\M_d$ does not contain
all locally maximally mixed states
\cite{baumgartnerHiesmayr}.\\
Depending on the focus of the analysis, several subsets and families of states
are of interest and have been discussed in literature~\cite{baumgartnerHiesmayr2,BaeQuasipure}. For
the entanglement analysis, the following families are
especially relevant: \\
\newline
\textbf{Enclosure polytope} \\
The enclosure polytope is defined as
\begin{gather}
\mathcal{E}_d \equiv \lbrace \rho = \sumkl c_{k,l} P_{k,l}~ |~
\sumkl c_{k,l} = 1, c_{k,l} \in [0, \frac{1}{d}]  \rbrace
\end{gather}
It was shown \cite{baumgartner1} that all states that lie outside of the
enclosure polytope are necessarily entangled. Since they can be detected by the
PPT or Peres-Horodecki criterion, they can be distilled by local operation and
classical communication (LOCC), or equivalently, are free and not bound
entangled \cite{distillation}.
\\
\newline
\textbf{Kernel polytope} \\
The kernel polytope is another geometric object that allows to determine the
entanglement class of a given state represented by its coordinates in $\M_d$. It
is defined as convex mixture of certain separable states, named ``line states''
$\rho_{\alpha}$ \cite{baumgartner1}:
\begin{gather}
\mathcal{K}_d \equiv \lbrace \rho = \sum_\alpha \lambda_{\alpha} \rho_{\alpha}~ |~
\lambda_\alpha \geq 0, \sum_\alpha \lambda_\alpha = 1 \rbrace
\end{gather}
The line states $\rho_\alpha$ are related to cyclic subgroups (or more general
sublattices in higher dimensions) of the linear ring structure induced by the
Weyl operators. As those are separable states, each state in
the $\mathcal{K}_d$ is separable by construction. \\
\newline \textbf{Family A} \\
Family A is defined by a mixture of three Bell states and the maximally mixed
state
$\rho_{mm} = \frac{1}{d^2} \sumkl \Pkl = \frac{1}{d^2} \sum_{i,j = 0}^{d-1}
\ketbra{ij}{ij}$ for $d=3$. The three Bell states are required to be on a phase
space line~\cite{baumgartner1}, i.e. one Bell state is chosen and the other two
are generated by application of one chosen Weyl transformation. Choosing without
loss of generality the Weyl transformation to be
$ W_{0,1} \otimes \mathbb{1}_d$, the states of Family A can explicitly be
written as
\begin{eqnarray}
  \label{famA}
\mathcal{F}_A \equiv \lbrace \rho_A &=& \alpha P_{00} + \beta P_{01} + \gamma
P_{02} + (1-\alpha -\beta - \gamma) \rho_{mm}\nonumber\\
&&|~ \rho_A \in M_3 \rbrace
\end{eqnarray}
Leveraging the high symmetry of such states, an optimal entanglement witness was
found~\cite{baumgartner1} and later related to the quasi-pure approximation of the concurrence
criterion~\cite{BaeQuasipure}. It is therefore possible to determine the
entanglement class of all states in this family.

\subsection{Numerical generation of states}
\label{sec:numer-gener-stat}
\begin{figure}
  \centering
  \includegraphics[width=0.5\textwidth]{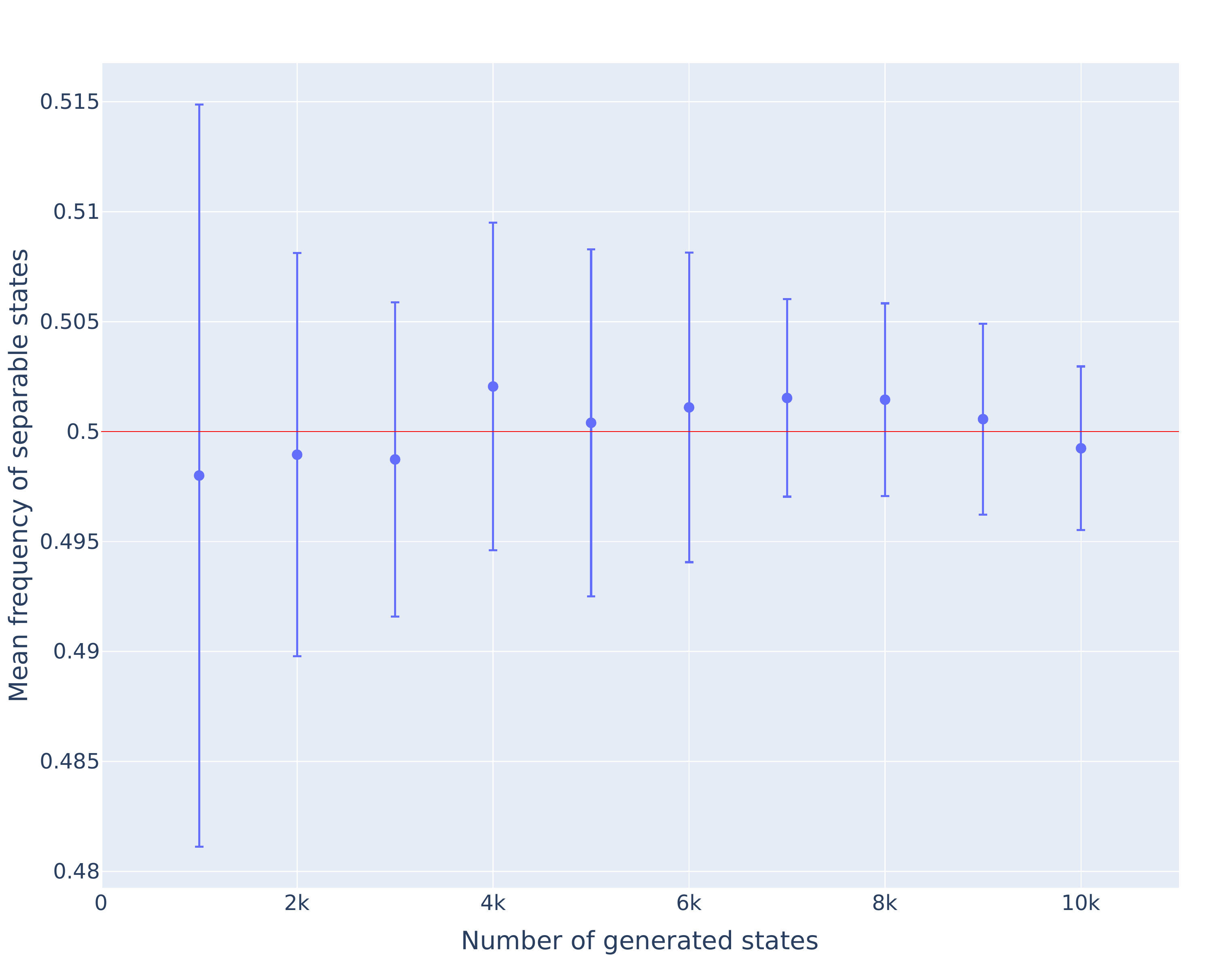}
  \caption{Relative frequencies of separable states in $M_2$. For each sample
    size, $10$ sets of randomly generated states are generated. The figure shows
    the empirical mean and standard deviation for the frequencies of separable
    states.}
    \label{d2volumes}
  \end{figure}

The representation of the system as $(d^2-1)$-dimensional simplex with the $d^2$
Bell states as vertices and the probabilities as ``baryocentric coordinates''
allows the efficient geometric representation of states as points in the
simplex. In this representation, states for the whole simplex, subsets of finite
volume or hyper-planes can be equivalently described and generated as points in
$d^2$-dimensional Euclidean space. This can be achieved by, e.g., random
sampling according to a certain distribution or by some deterministic
discretization procedure. In the following, two ways that allow the estimation
of relative volumes of separable, bound and free entangled states are shortly described.\\
\newline \textbf{Uniform random sampling of states} \\
The linear, real (sub-)spaces presented above allow uniform sampling of points
within and therefore uniformly distributed states can be generated in $\M_d$, $\mathcal{E}_d$
or $\mathcal{F}_A$. Classification of these states allows a probabilistic estimation of
relative volumes of the entanglement classes by using the relative
number/frequency of classified states as estimator. As the number of required
states for a valid estimation only depends of the relative volumes of the
classes and not on the dimension, this method of state creation is favorable for
volume estimation in higher dimensions. Another advantage of random sampling
compared to some deterministic procedure is that the expected relative number of
states in a given class only depends on the volume and not on the (unknown)
specific geometric shape of the set of states within that class.
\\
Let us demonstrate the validity of this method to estimate the relative volumes
of entanglement classes for bipartite qubits (i.e. $d=2$). In this case, the
separability problem can be solved analytically. It has been shown that for
$d=2$, all entangled states are free entangled and can be detected with the PPT
criterion \cite{peres,horodeckiCriterion} and therefore $\M_2$ contains no bound
entanglement. It is also known that the kernel polytope $\mathcal{K}_2$ contains
all separable states and that the relative volumes of both classes in $\M_2$ are
exactly $50 \%$. Figure~\ref{d2volumes} shows the relative frequency of
separable states for uniform samples in $\M_2$. For different sample sizes, the
empirical mean and standard deviation of $10$ runs are presented. One observes
that the relative frequency converges with growing number of states to the known
equal sized relative volumes of $0.5$ with increasing precision.  Given a random
sample of size $N$, the number of states within a certain class of relative
volume $p$ should be distributed according to the binomial distribution. The
expected number of states in that class is then $N \cdot p$ with standard
deviation of $\sqrt{N p (1-p)}$. In case of $\M_2$, the probability that
a generated state is PPT and in this special case also separable is $p=0.5$. \\

  \begin{table}
    \begin{tabular}{|l|l|}
      \hline
      \textbf{Sample set}
      & \textbf{\# PPT states}
      \\
      \hline
      1 & 4869 \\
      \hline
      2 & 5021  \\
      \hline
      3 & 4901 \\
      \hline
      4 & 5019  \\
      \hline
      5 & 4982 \\
      \hline
      6 & 4996  \\
      \hline
      7 & 5014 \\
      \hline
      8 & 5022  \\
      \hline
      9 & 4969  \\
      \hline
      10 & 5079 \\
      \hline
    \end{tabular}
    \caption{Number of PPT states in sample sets of size 10000 }
    \label{tab:binom-sample}
  \end{table}

Table~\ref{tab:binom-sample} shows ten sample sets containing
  $10000$ states each and the number of PPT states within. The
  empirical mean is $4987.2$ with empirical standard deviation of $61.7$. These
  results are in agreement with the expected mean of $5000$ and standard
  deviation of $50$ according to the binomial distribution. For the exemplary
  analysis below we use a sample size of $10000$ states. The determined
  empirical values imply that the obtained numbers for the size of relative

\begin{figure*}
  (a)\includegraphics[width=0.45\textwidth]{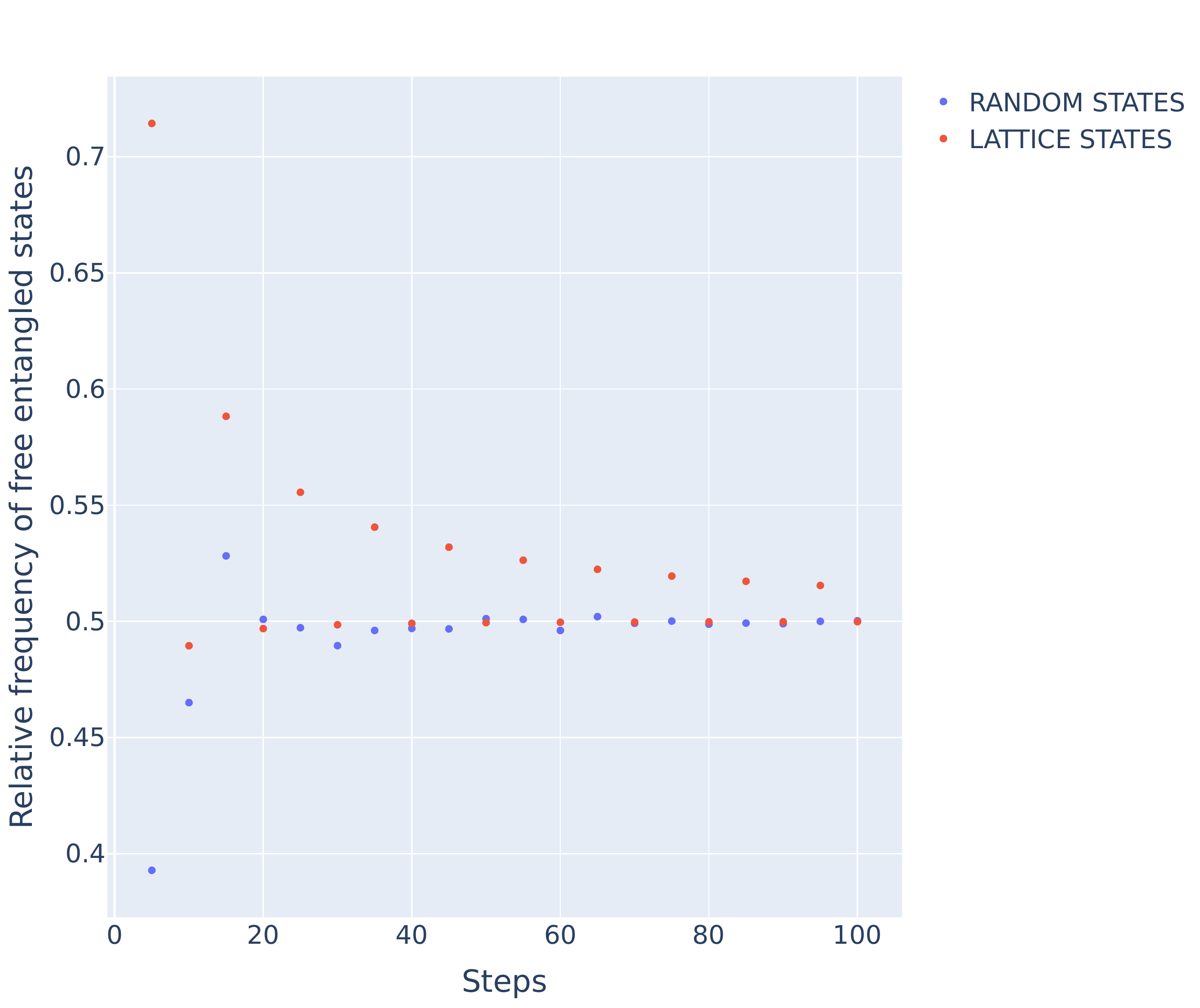}
  (b)\includegraphics[width=0.45\textwidth]{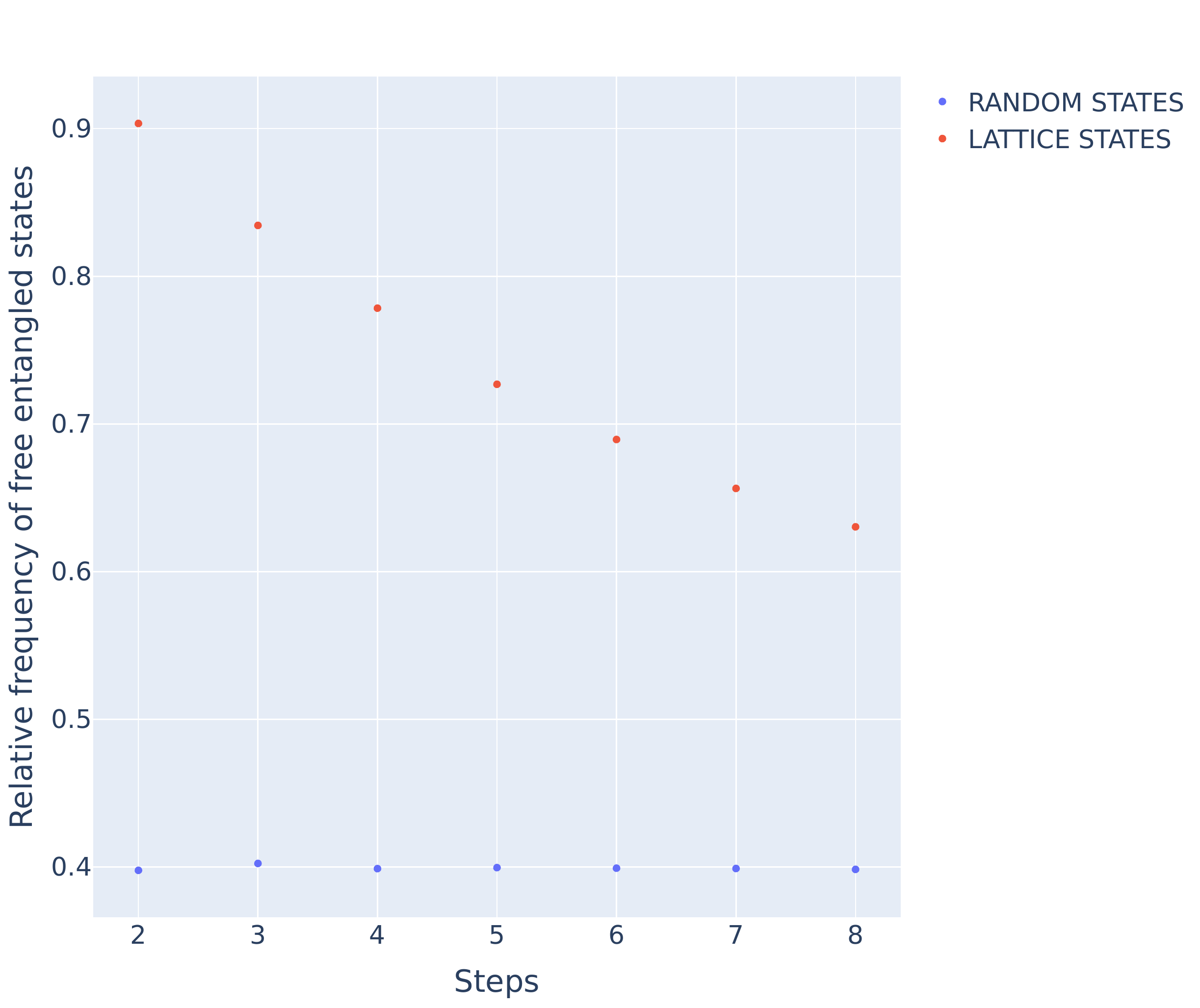}
\caption{Relative frequency of free entangled states generated on a lattice
  (red) and by random sampling (blue) in dependence of the number of
  steps for $\mathcal{M}_2$ for $d=2$ (left) and $\mathcal{E}_3$ for $d=3$ (right)}
  \label{convergence-lattice-rand}
\end{figure*}

\textbf{States on an equidistant lattice} \\
One way to discretize the space of interest is using a constant increment in all
dimensions. Given $k \leq (d^2-1)$ independent parameters, a certain range $R$
of values they can assume and a number of steps $s$ to divide the range in, one
obtains $(s+1)^k$ lattice points. Generating states on a fixed lattice offers
the advantage that these states have many symmetries, allowing to reduce the
number of states to explicitly classify. On the other hand, generating the full
lattice is quickly too computationally expensive with growing dimension, due to
the exponential dependency of the number of lattice points given a certain
number of steps. This makes this discretization computationally inefficient in
higher dimensions, because the step size needs to be small enough to capture the
entanglement structure. Another disadvantage of this set of states is that they
are only approximately uniformly distributed for small increments. One example
for that is the over-representation of states in certain areas depending on the
chosen discretization. As a consequence, classifying the states and counting the
relative occurrences as estimator for the relative volumes of the entanglement
classes is unbiased only in the limit of large $s$.  Figure~\ref{convergence-lattice-rand} compares the relative frequency of states that
are entangled according to the PPT criterion for random samples and states on a
fixed lattice. For $d=2$, the frequencies are relative to the whole simplex
$\M_2$, while for $d=3$ the enclosure polytope $\mathcal{E}_3$ is used. The
ranges of the coordinates are divided in a certain amount of steps to generate the
states on a lattice. For comparison, an equal number of random states is
generated and analyzed.  In Figure~\ref{convergence-lattice-rand} (left) one
observes that the randomly generated states quickly converge to the known
relative volume of $0.5$. The lattice states show a strong dependence on the
number of steps. While for an even number of steps the relative frequency
approaches the volume from below, for odd numbers the convergence is slower and
from above. Figure~\ref{convergence-lattice-rand} (right) shows that already for
$d=3$, the number of lattice states that can be generated by standard
computational means is not high enough to avoid a significant bias compared to
the quickly converging randomly generated states.

\subsection{Generation of symmetry transformations conserving the entanglement property}
\label{sec:gener-symm}
The phase space structure of $\M_d$ implies a set of ''entanglement-conserving''
symmetry transformations related to point transformations and translations in
the discrete phase space induced by Weyl operators \cite{baumgartnerHiesmayr,
  baumgartner1}. They act as permutations of the generating Bell projectors, or
equivalently, as permutations of the coordinates in $\M_d$ and are generated by
elementary transformations, i.e. generators of the involved symmetry
group. These generators can be identified with translation, momentum inversion,
quarter rotation and vertical shear in the induced discrete phase space. It was
shown that these symmetries conserve entanglement in $\M_d$. A simple argument
presented as Lemma \ref{symConserveEclass} in the Appendix also shows that the
entanglement class is conserved, meaning that the classification of free, bound
and separable states is not changed by applying any element of this group. The
generators and their actions can be represented as specific permutations,
allowing to generate all elements by succeeding application of several symmetry
transformations. Since the phase space contains a finite number of elements,
also the number of permutations is finite. This allows to determine all
symmetries generated by the above generators in arbitrary
dimension. The number of symmetries that can be generated by
those generators grows quickly with the dimension of the subsystems. For $d=2$,
$24$ such symmetries exist, while for $d=3$ already $432$ symmetries can
be found. Leveraging these symmetries is crucial to the analysis of entanglement
in $\M_d$, significantly improving the differentiation of bound entangled from
separable states.

\subsection{Sufficient criteria for entanglement}
\label{sec:suff-crit-entangl}
The problem to decide whether a given state is separable or entangled is
generally a NP-hard problem \cite{nphard}. No general solution by
polynomial in time algorithms is known and it is often denoted the ``separability
problem''. Likewise, there is no efficient method to solve the separability
problem for the mixtures of maximally entangled states $\M_d$ in general
dimension $d$. However, several sufficient criteria to detect entanglement are
known and can be used for entanglement classification. The most effective ones
for states in $\M_d$ are shortly stated below and enumerated by E1, E2,\dots. \\
\\
\textbf{E1: PPT criterion} \\
The ``Positive Partial Transpose (PPT)'' criterion \cite{peres}, also named
``Peres-Horodecki'' criterion states that if the partial transpose of a
bipartite state has at least one negative eigenvalue (in this case the state is
said to be ``NPT''), it is entangled. The partial transpose $\Gamma$ acts on the
basis states of a bipartite state as
$(\ketbra{i}{j} \otimes \ketbra{k}{l})^{\Gamma} \equiv \ketbra{i}{j} \otimes
\ketbra{l}{k}$. It can be calculated efficiently, but for $d \geq 3$ it is only
sufficient, not necessary, for entanglement. In this case, also states that have
a positive partial transpose can be entangled. Only entangled NPT-states can be
used for entanglement distillation are therefore denoted as free entangled.\\ \\
\textbf{E2: Realignment criterion} \\
The realignment criterion \cite{realignment} is structurally similar to the PPT
criterion. The realignment operation $R$ acts as
$(\ketbra{i}{j} \otimes \ketbra{k}{l})_R \equiv \ketbra{i}{k} \otimes
\ketbra{j}{l}$. The criterion states that if the sum of singular values of the
realigned state $\sigma_R$ are larger than $1$, i.e. if
$\tr \sqrt{\sigma_R^{\dagger} \sigma_R} > 1$, then $\sigma$ is entangled. Like
the PPT criterion, it can also be computed efficiently, but is only sufficient
for entanglement. It is neither stronger or weaker that the PPT criterion, so it
can
detect PPT entangled states, but generally does not detect all NPT-states. \\ \\
\textbf{E3: Quasi-pure concurrence criterion} \\
The quasi-pure approximation $C_{qp}$ of the concurrence
\cite{woottersConcurrence} provides another sufficient entanglement
criterion. States for which the concurrence is positive are entangled, but for
$d>2$ the concurrence allows only numerical estimates for mixed states. The
quasi-pure approximation \cite{BaeQuasipure} provides an easy to compute lower
bound for the concurrence and can therefore be used for entanglement
detection. Considering states of $\M_d$, an explicit form for the approximation
can be derived. A state $\rho = \sumkl c_{k,l}P_{k,l} \in \M_d$ is entangled
according to the quasi-pure concurrence criterion if
$C_{qp}(\rho) = \max(0, S_{nm} - \sum_{(k,l) \neq (n,m)} S_{k,l})>0$ where
$(n,m)$ is a multi-index of the coordinate of the largest mixing probability $\lbrace c_{k,l} \rbrace $
and $S_{k,l}$ are explicitly given by \cite{BaeQuasipure}:
\begin{widetext}
\begin{gather}
S_{k,l} = \sqrt{
  \frac{d}{2(d-1)} c_{k,l}
  [(1-\frac{2}{d}) c_{n,m} \delta_{k,n} \delta_{l,m}
  + \frac{1}{d^2} c_{(2n-k)mod~d,(2m-l)mod~d}]
}
\end{gather}

\textbf{E4: MUB criterion} \\
In a Hilbert space of dimension $d$, a set of orthonormal bases $\lbrace
B_k \rbrace$ and $B_k = \lbrace \ket{i_k} ~|~ i = 0,\ldots, (d-1) \rbrace$
is called ``mutually unbiased bases (MUB)'' if $\forall k \neq l$:
\begin{gather}
  |\braket{i_k}{j_l}|^2 = \frac{1}{d} ~~~ \forall i,j = 0, \ldots, (d-1)
\end{gather}

Given $m$ MUBs, it was shown \cite{hiesmayrLoeffler, SpenglerMUB} that the sum
of all ``mutual predictabilities`` $I_m$ is bounded from above for separable states
$\rho_s$:
\begin{gather}
  I_m(\rho_s) =
  \sum_{k=1}^m C_k =
  \sum_{k=1}^m \sum_{i=0}^{d-1} \bra{i_k}\otimes \bra{i_k} \rho_s \ket{i_k}
\otimes \ket{i_k} \leq 1 + \frac{m-1}{d}
\end{gather}
\end{widetext}
At most $d+1$ MUBs exist \cite{woottersMUB, BandyopadhyayMUB} in which case
$ I_{d+1}(\rho_s) \leq 2 $ for all separable states $\rho_s$. Conversely, if an
unclassified state exceeds this upper bound, it is entangled.  In order to
detect bound entanglement in $d=3$, the $C_k$ can be modified to the following
form:
\begin{gather}
  \label{eq:2}
  C_1 = \sum_{i=0}^{2} \bra{i_1}\otimes \bra{(i_1+2\pmod 3)^*} \nonumber\\ \qquad\quad\qquad\qquad\rho_s
  \ket{i_1}\otimes \ket{i_1+2\pmod 3^*}, \nonumber\\
  C_k =   \sum_{i=0}^{2} \bra{i_k}\otimes \bra{i_k^*} \rho_s \ket{i_k} \otimes
  \ket{i_k^*},~ k=2,3,4
\end{gather}
Here, $i_k^*$ denotes the complex conjugate vector element.
In this form, like the realignment and quasi-pure concurrence criteria, the MUB
criterion allows the detection of PPT entangled states, which was also experimentally
demonstrated for entangled photons \cite{hiesmayrLoeffler}.
\\ \\
\textbf{E5: Numerically generated entanglement witnesses}\\
Leveraging the fact that separable states form a convex set, entanglement
witnesses \cite{EWs} (``EWs'') are an important tool to detect entangled
states. An EW $W$ is an observable which implies an upper bound $U$ and, as
recently shown \cite{EW20}, also a lower, mostly nontrivial, bound $L$
($U,L \in \mathbb{R}$), for separable states $\rho_s$:
\begin{gather}
  \label{eq:1}
  L \leq \tr[\rho_s W] \leq U
\end{gather}
If a state $\rho$ suffices $\tr[\rho W] \notin [L,U]$ for some $W$, then it is
entangled and said to be ``detected by $W$''. Originally EWs were introduced
considering the upper bound $U=0$ only. It has been shown \cite{baumgartner1}
that for each entangled state $\rho \in \M_d$ there exists an EW $W_\rho$ such
that $W_\rho$ detects $\rho$. In this sense, EWs are universal, because each
entangled state can be detected by an, unfortunately generally unknown, EW.  If
$\rho \in \M_d$, it suffices to consider EWs of the form
$W = \sumkl \kappa_{k,l} \Pkl$ with $\kappa_{k,l} \in [-1,1]$ to detect all
entangled states \cite{baumgartner1}. In this case
$\rho = \sumkl c_{k,l}\Pkl \in \M_d$ and we have
$\tr[\rho W]= \sumkl c_{k,l} \kappa_{k,l} \equiv c \cdot \kappa$ where the dot
indicates the standard scalar product of the $d^2$-dimensional vectors $c$ and
$\kappa$ collecting the coefficients $c_{k,l}$ and $\kappa_{k,l}$. Thus, in the
geometric representation of $\M_d$, any EW defines two $(d^2-1)$-dimensional
hyper-planes via $c_L \cdot \kappa = L$ and $c_U \cdot \kappa = U$ and
corresponding halfspaces. Any state that is represented by a point outside of
the intersection of these halfspaces is detected as entangled. The difficulty of
using EWs lies of course in determining the bounds $L$ and $U$ for the set of
separable states. An efficient parameterization of unitaries
\cite{spenglerHuberHiesmayr2} is used to numerically determine the bounds for
our case (see Appendix 2 for more details).
\subsection{Sufficient criteria for separability}
\label{sec:suff-crit-separ}
Since the presented criteria for entanglement are only sufficient and generally
not necessary, it is highly desirable to develop methods to directly classify
separable states in as well. Few analytical sufficient criteria for separability
exist, but the geometric characterization and properties of $\M_d$ allow the
development of an effective procedure to identify states of $SEP \cap
\M_d$. Here we present a new method based on the extension of the convex hull of
known separable states (S1) and a criterion with close relation to the states of
$\M_d$ (S2).\\ \\
\textbf{S1: Extended kernel criterion}\\
Given a finite set of known separable states and their representation in $\M_d$,
a convex polytope can be constructed by building the convex hull of those
vertices. All states of $\M_d$ with coordinates within this polytope are
separable as well. The problem to decide whether a given Bell diagonal state is
a convex combination of known existing separable states is then equivalent to
the standard problem of linear programming to decide whether a given point lies
within a convex polytope. Several numerical implementations to solve that
problem exist, e.g. Ref.\cite{lazysets}, which was used for this work. Obviously, in
order to check an unknown state $\rho \in \M_d$ for separability, a set of known
separable states as vertices to build the convex hull are required. Since the
effectivity of the check depends on the volume of the spanned polytope, it is
advantageous to use vertices that cover uniformly distributed spatial angles and
are as close to the boundary of the convex set of separable states as
possible. The line states $\rho_\alpha$ that build the kernel polytope $\mathcal{K}_d$
meet those requirements, as they are known to be on the surface of $SEP$
\cite{baumgartner1}. However, more separable states/vertices are needed to cover
approximately all separable states in $\M_d$. For the results presented in this
work, additional separable states were generated and multiplied by using the
entanglement-class-conserving symmetries to
generate more vertices to extend the separable kernel.\\
\\
\textbf{S2: Weyl/Spin representation criterion}\\
The Weyl operators $W_{k,l}$ satisfy useful relations to simplify required
calculations for states of $\M_d$. Two of those are:
\begin{gather}
  W_{k_1,k_2}W_{l_1,l_2} = w^{k_2 l_1}W_{k_1+l_1, k_2+l_2}  \\
  W_{k_1,k_2}^\dagger = w^{k_1 k_2}W_{-k_1, -k_2} = W_{k_1, k_2}^{-1}
\end{gather}
It follows that the Weyl operators form an orthogonal family of unitary matrices
with respect to the trace norm $(A.|B) \equiv \tr[A^\dagger B]$ and as such is a
basis for the space of $d \cross d$ matrices. Any density matrix $\sigma$ can
then be represented as $ \sigma = \frac{1}{d} \sumkl s_{k,l} W_{k,l}$. The
coefficients of this Weyl representations are
$s_{k,l} = \tr[W_{k,l}^\dagger \sigma]$. Accordingly,
$W_{\mu, \nu} \equiv W_{\mu_1, \nu_1} \otimes W_{\mu_2, \nu_2}$ can be used to
represent a bipartite state $\rho$ with coefficients $s_{\mu, \nu}$. A
sufficient criterion for separability was derived \cite{weylRepCrit} that is
named ``Weyl'' or ``Spin representation criterion'' here: If
$\sum_{\mu, \nu} |s_{\mu, \nu}| \leq 2$, where $|s_{\mu, \nu}|$ are the
coefficients of the Weyl representation of the bipartite state $\rho$, then
$\rho$ is separable.
\subsection{Symmetry analysis}
Leveraging the rich symmetries of $\M_d$ is another crucial factor to detect both
entanglement or separability with the presented methods. Using Lemma
\ref{symConserveEclass} about the conservation of the entanglement classes and
the generation of all elements of this symmetry group greatly improves the
detection capabilities of available detectors. Given an unknown state, all
symmetric states can generated as by applying the according transformation for
all generated symmetries. This set is then analyzed with respect to the
available criteria. If entanglement class is determined for one state, Lemma
\ref{symConserveEclass} ensures that all states of the set of symmetric states
are known to be of the same class as well.

\section{Results}
We determine the share of separable, bound and free entangled states with high
accuracy using the methods presented above. After summarizing arguments for the
generation of states via random sampling, two exemplary analyses are presented.
First, states of the subset $\mathcal{F}_A$ are analyzed before the main results
of the entanglement classification for the enclosure polytope $\mathcal{E}_3$
for the $d=3$ are presented. The analysis of $\mathcal{F}_A$ serves as
validation and comparison of applied methods, because here the borders between
separable, bound entangled and free entangled states are analytically known
\cite{baumgartnerHiesmayr,BaeQuasipure}. Since we know that all states outside
of the enclosure polytope $\mathcal{E}_d$ are free entangled, the analysis of
$\mathcal{E}_3$ suffices to know the entanglement classification of the whole
simplex $\M_3$. Entanglement and separability criteria are compared for their
effectiveness and their relations are discussed.  The states are classified
according to the labels ``SEP'' (separable states), ``BOUND'' (bound/PPT
entangled states), ``FREE'' (free/NPT entangled states) and ``PPT-UNKNOWN''
(PPT states that could not be shown to be separable or bound entangled). For
that purpose, more than $16,000$ entanglement witnesses have been generated
numerically. Additionally, the kernel of separable states has been extended by
new separable states leveraging the generated symmetries to generate new vertex
states for the convex hull. The symmetries have also been used for all methods
to further increase the detection capabilities (see section ``Symmetry
analysis'' above).

\subsection{Generation of states}
The generation of uniformly distributed random states offers significant
advantages compared to states on a fixed lattice for estimating the relative
volumes of entanglement classes in subsets of $\M_d$. \\
First, the relative frequencies of states in a certain class provides an
unbiased estimator for the relative volume of that class if the states are
randomly sampled but not if they are generated on a lattice with fixed
increment. As shown in Figure~\ref{convergence-lattice-rand} (right), the
relative frequencies of randomly generated states converge quickly with
increasing sample size, while the convergence of lattice states is slower for
$d=3$ so that a significant bias
remains for a number of states of the same magnitude.\\
Second, the expected relative frequencies of random states only depend on the
relative volume of the classes and not on their specific geometric shape. A
fixed lattice, however, may not distribute its states equally between the
classes (according to their volume) as can be observed in Figure~\ref{convergence-lattice-rand} (left). For $d=2$, the relative frequencies
depend strongly on the increment of the lattice and show different convergence
rates and directions for even and odd number of lattice points in each dimension
of $\M_2$. Since the geometric shape of entanglement classes in $\M_d$ is
generally unknown for $d \geq 3$, it is not possible to make quantitative
statements about the estimation and its error. Here, the uniform generation of
random states is advantageous as well, since the probability of finding a random
state to be in a given entanglement class is equal to its relative volume. The
expected number of states in that class and the standard deviation thereof is
then given by the binomial distribution, which was demonstrated for $d=2$ in
Table \ref{tab:binom-sample}. Even if the exact size of the classes (and
therefore the distribution) is not known \textit{a priori}, the obtained
estimations can be used to approximate the expected deviation of the true volume
and the estimation.\\
Finally, the validity of this method allows the extension for higher
dimensions. While already for $d=3$ it is not possible to generate enough states
based on a lattice to estimate the relative volumes correctly, the problem will be
even larger for $d > 3$ due to the exponential growth of the Hilbert space. For
randomly generated states, however, the estimation and its variance depend only
on the size of the classes and not on the dimension. This implies that the
sample size does not need to be increased for an estimation of the same quality
in higher dimensions.

\subsection{Entanglement classification for $\mathcal{F}_A$}
\label{sec:classication-F_A}
To analyze Family A, the parameters $\alpha, \beta, \gamma$ of equation
(\ref{famA}) are uniformly sampled within the range $[-1,1]$ and the
corresponding states are checked for positivity. This way, $10,000$ uniformly
distributed states of $\mathcal{F}_A$ are generated and analyzed.

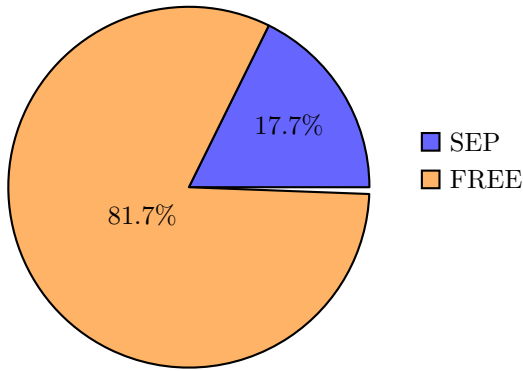
\begin{figure}[h]
\begin{minipage}{0.4\textwidth}
  \begin{table}[H]
    \centering
    \begin{tabular}{|l|r|r|}
      \hline
      \textbf{  Entanglement Class  } &  \textbf{  \# states  } & \textbf{  \#
                                                                  relative to
                                                                  $\mathcal{F}_A$  }  \\
      \hline
      Total $\mathcal{F}_A$  & $10000$ & $100\%$ \\
      \hline
      SEP & $1774$ & $17.7 \%$ \\
      \hline
      BOUND & $7$ & $0.1 \%$ \\
      \hline
      FREE & $8171$ & $81.7 \%$ \\
      \hline
      PPT-UNKNOWN & $48$ & $0.5 \%$ \\
      \hline
    \end{tabular}
  \end{table}
\end{minipage}
\hspace{6em}
\begin{minipage}{0.4\textwidth}
  \begin{tikzpicture}[scale=.8]
    \pie[text=legend, color={blue!60, orange!60}]{
      17.7/SEP,
      81.7/FREE
    }
  \end{tikzpicture}
\end{minipage}
\caption{Entanglement classes and their relative volumes in $\mathcal{F}_A$}
\label{tab:classification-FA}
\end{figure}

Figure~\ref{tab:classification-FA} shows that more than $99.5 \%$ of the states
could be classified. The majority of states in $\mathcal{F}_A$ are free
entangled ($81.7\%$), $17.7 \%$ are separable and only $0.1 \%$ of all states
are found to be bound entangled. For the remaining $0.5 \%$, the states are
known to have a positive partial transpose, but none of the criteria detected
separability or
bound entanglement.\\
As mentioned above, the criterion $E3$ was shown to be optimal for this
family. This is also reflected in our results in table
\ref{tab:bound-sep-detectors-FA}, showing that only E3 detects bound entangled
states. It can therefore be concluded, that the PPT-UNKNOWN states should be
separable as well. Concerning separable states, both criteria S1 and S2 detect a
significant number of states in SEP. However S1 is clearly stronger than S2,
because all states detected by S2 are also detected by S1.

\begin{table*}
\centering
\begin{tabular}{|l|l|r|r|}
  \hline
  \textbf{Entanglement Class} &  \textbf{Criterion} & \textbf{\# detected} &
                                                      \textbf{\# relative to class}  \\
  \hline
  SEP & S1 & 1774  & $100\%$ \\
  \hline
  SEP & S2 & 457  & $25.8\%$ \\
  \hline
  BOUND & E2 & 0  & $0 \%$ \\
  \hline
  BOUND  & E3 & 7  &$100 \%$ \\
  \hline
  BOUND & E4 & 0  &$0 \%$ \\
  \hline
  BOUND & E5 & 0  &$0 \%$ \\
  \hline
\end{tabular}
\caption{BOUND and SEP detectors in $\mathcal{F}_A$ and their relative number in class}
\label{tab:bound-sep-detectors-FA}
\end{table*}

\subsection{Entanglement classification for $\mathcal{E}_3$}
\label{sec:classication-d=3}
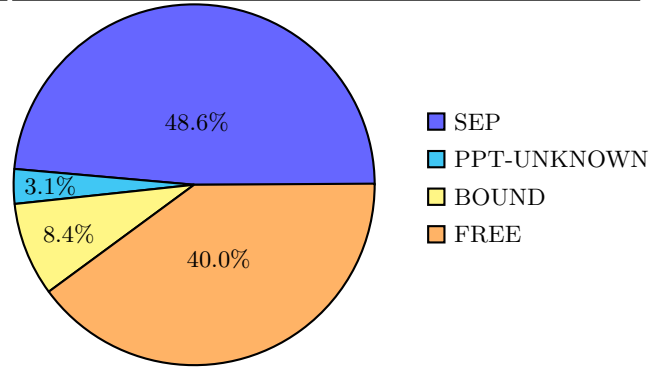
\begin{figure}[H]
\begin{minipage}{0.4\textwidth}
  \begin{table}[H]
    \centering
    \begin{tabular}{|l|r|r|}
      \hline
      \textbf{  Entanglement Class  } &  \textbf{  \# states  } & \textbf{  \#
                                                                  relative to
                                                                  $\mathcal{E}_3$  }  \\
      \hline
      Total $\mathcal{E}_3$  & $10000$ & $100\%$ \\
      \hline
      SEP & $4858$ & $48.6 \%$ \\
      \hline
      BOUND & $835$ & $8.4 \%$ \\
      \hline
      FREE & $3999$ & $40.0 \%$ \\
      \hline
      PPT-UNKNOWN & $308$ & $3.1 \%$ \\
      \hline
    \end{tabular}
  \end{table}
\end{minipage}
\hspace{6em}
\begin{minipage}{0.4\textwidth}
  \begin{tikzpicture}[scale=.8]
    \pie[text=legend]{
      48.6/SEP,
      3.1/PPT-UNKNOWN,
      8.4/BOUND,
      40.0/FREE
    }
  \end{tikzpicture}
\end{minipage}
\caption{Entanglement classes and their relative volumes in $\mathcal{E}_3$}
\label{tab:classification}
\end{figure}

The main results of this contribution is the determination of the entanglement
class for arbitrary mixtures of Bell diagonal states of which $\mathcal{F}_A$ is
a small subset. We restrict the analysis to the enclosure polytope
$\mathcal{E}_3$ since all other states of $\mathcal{M}_3$ are
known to be free entangled according to the PPT criterion.\\
For this analysis, $10,000$ uniformly distributed random states are generated and
analyzed. The results are summarized in Figure \ref{tab:classification},
showing the number of states and the relative volume for each class. Almost half
of the states ($48.6 \%$) are shown to be separable states via the separability
criteria. $48.4 \%$ are entangled, of which $40.0 \%$ are found to be free
entangled according to the PPT criterion as already expected with the analysis
of Figure \ref{convergence-lattice-rand}. $8.4 \%$ of the states are bound
entangled and only $3.1\%$ of the states are known to be PPT-states, but none of
the criteria allowed the certain detection of entanglement or separability, so
the entanglement class remains unknown. To confirm the classification, we
applied the detection criteria for the other classes also to already classified
states, which all failed.
\\

\ \\
These results extend the previous related investigation \cite{hiesmayr1} in
several ways. First, the presented methods allow a more exact estimation of the
relative volumes of the entanglement classes in $\M_3$. As argued above, the
random generation of states provides an unbiased estimator for the relative
volumes. In the cited analysis, the states were generated on a lattice with
constant increment, but as shown in Figure \ref{convergence-lattice-rand} the
increment is not small enough to ignore the systematic bias due to the fixed
lattice. In the cited analysis, higher values for the relative volumes of FREE
and BOUND entangled states are given, while less states are classifies as
SEP. Figure \ref{convergence-lattice-rand} indicates that the lattice
construction generates a higher density of states in areas in which FREE and
BOUND entangled states dominate. Consequently, the generated states are not
uniformly distributed, but, as can be easily seen, have a higher density for
larger distances to the center of the kernel $\mathcal{K}_3$. This introduces a
shift in the relative frequencies of the entanglement classes of generated
states, since on average, separable states are closer to the center than
entangled states.  Second, using the new method S1 and S2 allows the direct
detection of separable states beyond the kernel polytope
$\mathcal{K}_3$. Together with the creation and consideration of all available
symmetries it is possible to significantly reduce the number of PPT-UNKNOWN
states.
\\
The main challenge in the classification is of course the differentiation of
separable and bound entangled states. For this reason we are strongly interested
in the detection capabilities of different criteria for these classes, which are
summarized in Table~\ref{tab:bound-sep-detectors}.

\begin{figure*}
  \centering
  \includegraphics[scale=0.45]{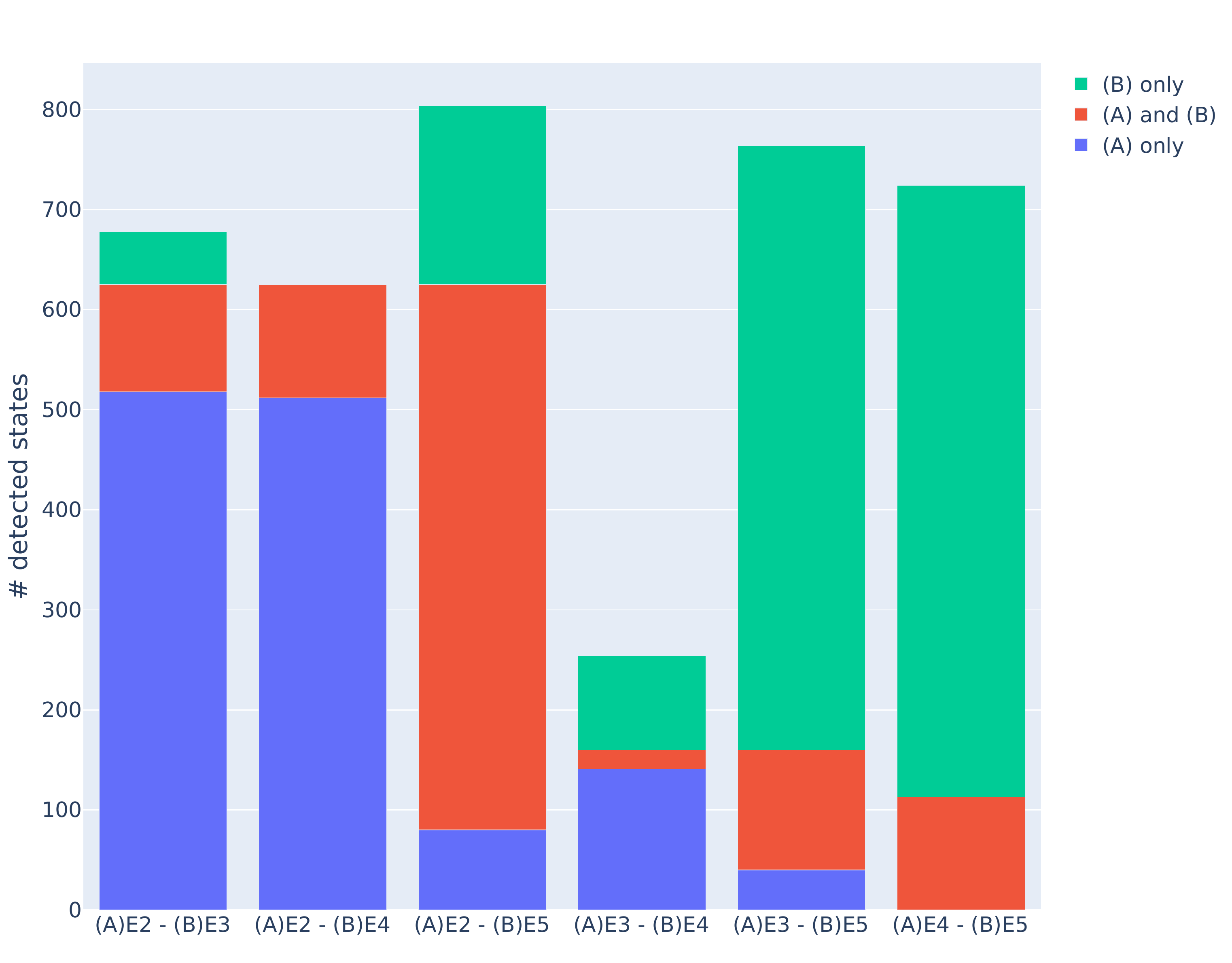}
  \caption{Pairwise comparison of number of exclusively (blue and green) and
    jointly (red) detected states}
  \label{inter}
\end{figure*}

\begin{table*}
\centering
\begin{tabular}{|l|l|r|r|}
  \hline
  \textbf{Entanglement Class} &  \textbf{Criterion} & \textbf{\# detected} &
                                                      \textbf{\# relative to class}  \\
  \hline
  SEP & S1 & 4858  & $100\%$ \\
  \hline
  SEP & S2 & 76  & $1.7\%$ \\
  \hline
  BOUND & E2 & 625  & $74.9 \%$ \\
  \hline
  BOUND  & E3 & 160  &$19.1 \%$ \\
  \hline
  BOUND & E4 & 113  &$13.5 \%$ \\
  \hline
  BOUND & E5 & 724  &$86.6 \%$ \\
  \hline
\end{tabular}
\caption{BOUND and SEP detectors and their relative number in class}\label{tab:bound-sep-detectors}
\end{table*}
\ \\
Although the analytical separability criterion S2 detects some states, the share
($1.7 \%)$ is very small and the criterion S1, leveraging the geometric
representation of the system and its symmetries, is clearly stronger. It detects
all states for which separability is implied by S2 and many additional
ones. Concerning the detection of bound entangled states, two criteria are
especially powerful: First the analytical criterion E2, which detects $74.9 \%$
of all determined bound entangled states and second E5, using the set
numerically generated EWs to detect $86.6 \%$ of the bound states. The later
especially leverages the symmetry analysis to detect additional states. E3 and
E4, although less effective than the other criteria, still detect a significant
portion ($19.1 \%$ and $13.5 \%$) of PPT entangled states. Note also that E3 is
the criterion that detects all bound entangled states for the family
$\mathcal{F}_A$.
\\
The question arises, whether some of the applied criteria to detect bound
entangled states are stronger than others for $E_3$. A criterion (A) is said to
stronger than criterion (B), if (A) detects all states detected by (B), as
well. Table~\ref{tab:pair-comp-bound} relates a pair of applied entanglement
criteria by comparing the number of exclusively detected bound entangled states
to the number of bound entangled states that were detected by
both criteria. The results are visualized in Figure \ref{inter}.\\
\begin{table*}
\centering
\begin{tabular}{|l|l|l|l|l|}
  \hline
  \textbf{Criterion (A)}
  & \textbf{\# detected (A)}
  &\textbf{Criterion (B)}
  &\textbf{\# detected (B)}
  &\textbf{\# detected (A) and (B)}
  \\
  \hline
  E2 & 625 & E3 & 160 & 107 \\
  \hline
  E2 & 625 & E4 & 113 & 113 \\
  \hline
  E2 & 625 & E5 & 724 & 545 \\
  \hline
  E3 & 160 & E4 & 113 & 19 \\
  \hline
  E3 & 160 & E5 & 724 & 120 \\
  \hline
  E4 & 113 & E5 & 724 & 113 \\
  \hline
\end{tabular}
\caption{Pairwise comparison by criterion of detected bound states}
\label{tab:pair-comp-bound}
\end{table*}

\ \\
The set of bound entangled states detected by the two most effective criteria,
E2 and E5, have a significant intersection, but each criterion also detects
states that the other does not. $53$ of $160$ states ($33\%$) that are detected
by E3 are not found by E2, showing that E3 is not really weaker than E2, but
seems to detect states of different areas of the geometrical polytope
$\mathcal{E}_3$. In this sense, the analytical criteria E2 and E3 complement
each other to a certain degree, as they together detect $81\%$ of all detected
bound entanglement. A similar, but slightly weaker conclusion can be made for
the relation of E3 and E5, where $25\%$ of the states found by E3 are not found
by E5. This is in contrast to the results of E4. Here, both E2 and E5 detect all
of the states found to be entangled according to E4, too. However, only $17\%$
of the states detected by E4 are also found by E3, showing again a structural
difference in the detection capabilities of the criteria E3 and E2. Also note
that more than two entanglement criteria are needed to detect all bound
entangled states.\\
In summary, the analytical criterion E2 and the numerically generated E5 are the
most effective detectors. Their detected states have a significant intersection,
but none is stronger than the other. In principle, all bound entangled states
can be detected by E5 with sufficiently many generated EWs. Thus, the states
detected by E2 and not by E5 correspond to states in the space between the
convex set of separable states and the enclosing hyper-planes of generated
linear EWs. E3 detects less states than E2, but has a high share of exclusively
detected states.  E4 can be considered as weaker than E2 and has a very small
overlap with E3, which further supports the hypothesis that E3 has a structural
difference compared to the other criteria.

\section{Discussion and Summary}
We investigated mixtures of maximally entangled, bipartite Bell states in three
dimensions. The states are locally maximally mixed, so all information is in the
correlation of the subsystems, while no information about the individual systems
is available. Depending on the mixing probabilities, the mixed state can be
separable or entangled. Curiously, also bound entangled states, i.e. states that
cannot be distilled by LOCC, occur, motivating a detailed analysis of this
system to study this exotic form of entanglement. \\
The investigated system allows a geometric representation in which its states
and their properties can be effectively geometrically analyzed. Leveraging these
properties, the NP-hard problem of determining whether a state is separable,
bound or free entangled can be almost completely solved for Bell diagonal
qutrits with new efficient methods. Three main aspects contribute to this
result: First, an efficient creation of states to estimate the relative volumes
of entanglement classes. Second, the numerical creation of symmetries and their
utilization for entanglement classification. Third, the development of
independent separability and entanglement detection criteria, which can be
applied efficiently to the to be classified states. A new collection of
computational methods has been implemented to realize these aspects. It has been
shown that the random sampling of states can be used to estimate the relative
volumes of entanglement classes without systematic bias and that is has
significant advantages compared to using states on a fixed
lattice.\cite{hiesmayr1} Further utilizing the generators of the considered
symmetries explicitly, all elements of the related group could be generated and
applied to improve applied criteria to analyze the entanglement structure. The
rapid growth of the number of distinct symmetry transformations with the
dimension suggests that our methods can also be applied to higher dimensions
despite the exponential growth of the Hilbert space.
\\
Leveraging an efficient parameterization of separable states and the generation
of symmetries allowed to use a sufficient criteria to directly detect separable
states as well as bound entangled states using numerically generated entanglement
witnesses. Several other well known entanglement criteria such as the PPT or
realignment criterion have been implemented and applied to a representative
sample of the system, allowing to compare their effectiveness.
\\
Two relevant subsets of bipartite states in $\M_d$ for $d=3$ were
analyzed. First family $\mathcal{F}_A$, which contains mixtures of three Bell
states and the maximally mixed state and whose symmetries allow to show that all
bound entangled states can be detected by an analytical witness
(E3).\cite{baumgartnerHiesmayr,BaeQuasipure} More than $99 \%$ of the states
could be classified including the detection of bound entanglement, but only
$0.1 \%$ are found to be of that class. The numerical implementation of E3
confirms its effectiveness, since all bound entangled states are detected by
this criterion. Interestingly, none of the other criteria detect any bound
entanglement for this family. The second analysis and main result of this work
is the entanglement classification of the enclosure polytope $\mathcal{E}_3$,
which is known to contain all PPT states. Again $10,000 $ uniformly distributed
states were generated as representative sample of the system of which $96.9\%$
could be classified. Only for $3.1\%$ it remains unknown whether the state is
separable of bound entangled. Almost half ($48.6\%$) are found to be certainly
separable, $40.0\%$ are free entangled and can therefore be distilled by LOCC
and at least $8.4\%$ are bound entangled. The developed numerical methods S1 and
E5 leveraging the special properties of the investigated system are the most
effective criteria to distinguish separable and bound entangled states. However,
also general analytical criteria detect significant shares of the bound
entangled states, especially the ``realignment criterion'' E2 is very effective
as well. All methods used to detect bound entanglement were investigated in
pairwise comparison. Although not being the most effective method in terms of
total number of detected states, the nonlinear criteria E3 (``quasi-pure
concurrence''), detects a significant amount of states that are not detected by
other criteria, especially the linear witness E4 and the collection of linear
witnesses E5. This indicates that the states detected by E3 and not by E4 and E5
relate to states close to the nonlinear surface of separable states.  It is
interesting that E3 is such a strong detector of bound entanglement in
$\mathcal{F}_A$, while other criteria like E2, which is a strong detector in
$\mathcal{E}_3$, fail for this specific subset.
\\
The developed methods and constructed related objects (i.e. generated entanglement
witnesses and extended kernel) can be repeatedly applied to new states of
interest in an efficient way. In this sense, the NP-hard problem of entanglement
classification is efficiently solved for this system with an accuracy of
approximately $94.9\%$, because any Bell diagonal PPT state can be efficiently
classified with high probability. Only for $5.1\%$ of the analyzed PPT states we
fail to classify it as separable of bound entangled. Moreover, we can deduce
from our studies that least $13.9\%$ of PPT Bell diagonal states are bound
entangled while $81.0\%$ are separable. The implemented framework can equivalently
be applied to any dimension $d$. It will be interesting to see the results in
higher dimensions $d=4$ and $d=5$ and what structural similarities and
differences in the classification of entanglement can be identified.\\
\\
\textbf{Acknowledgments:} B.C. Hiesmayr acknowledges gratefully the Austrian Science Fund (FWF-P26783).


\section{Appendix}
\label{appendix}

\subsection*{A1: Entanglement class conserving symmetries}
\label{sec:entangl-class-pres}
It was shown that the symmetries presented in Ref. \cite{baumgartnerHiesmayr}
for $d=3$ and in Ref. \cite{baumgartner1} for general dimension
conserve entanglement for states in $\M_d$, i.e. separable states are mapped to
separable states and entangled states to entangled states. However, no statement
is made about the possibility of free entangled states being mapped to bound
entangled states and vice versa. That this does not happen and that therefore
the symmetries conserve the entanglement class of separable, bound and free
states can be seen by the following argument.

\begin{mylemma}
  \label{symConserveEclass}
  The entanglement conserving symmetries of Ref. \cite{baumgartner1} conserve the
  entanglement class in $\M_d$.
  \begin{proof}
    Let $\rho \in \M_d$ and $S$ be the group of all entanglement conserving
    symmetries. Denote any element $s\in S, s: \M_d \rightarrow \M_d$ and the
    existing inverse by $s^{-1}$. Denote the set of states with positive partial
    transposition as $PPT$ and the complementary set as $NPT$ containing free
    entangled states.The generators of $S$ are represented either by local
    unitary transformations or the global complex conjugation. Both map $PPT$
    onto itself so
    \begin{gather}
      \label{eq13}
      s: PPT \cap \M_d \rightarrow PPT \cap \M_d.
    \end{gather}
    Now suppose $s:NPT \cap \M_d \rightarrow PPT \cap \M_d$ then
    $\mathbb{1} = s^{-1}s: NPT \cap \M_d \rightarrow PPT \cap \M_d \rightarrow PPT
    \cap \M_d$ is a contradiction. Here the first relation is the assumption and
    the second follows from (\ref{eq13}). So we also have:
    \begin{gather}
      \label{eq16}
      s:NPT \cap \M_d \rightarrow NPT \cap \M_d
    \end{gather}
    Denote the set of bound entangled states by $BOUND$. A similar argument
    shows that bound states are again mapped to bound states: By definition of
    entanglement conserving symmetries we have
    \begin{gather}
      \label{eq14}
          S: SEP \cap \M_d \rightarrow SEP \cap \M_d.
    \end{gather}
    Now suppose $s:BOUND \cap \M_d \rightarrow SEP \cap \M_d$ then
    $\mathbb{1} = s^{-1}s: BOUND \cap \M_d \rightarrow SEP \cap \M_d \rightarrow
    SEP \cap \M_d$ is a contradiction. Again the first relations is the
    assumption and the second from (\ref{eq14}). Together with (\ref{eq13})
    this shows that bound entangled states are mapped to entangled states
    which are also PPT:
    \begin{gather}
      \label{eq15}
          S: BOUND \cap \M_d \rightarrow BOUND \cap \M_d.
    \end{gather}
    (\ref{eq16}), (\ref{eq14}) and (\ref{eq15}) together show the invariance of
    the classes free entangled, separable and bound entangled under the
    entanglement conserving symmetries.
  \end{proof}
\end{mylemma}

\subsection*{A2: Parameterization of separable states to determine bounds for
  numeric EWs}
\label{sec:param-unit-stat}
An efficient parameterization of the set of separable states is required for
entanglement detection with numeric EWs (E5). Using an EW
$W$ requires the determination of its upper and lower bound for the set of
separable states $\rho_s \in SEP$: $L \leq \tr[\rho_s W] \leq U$. To determine
these bounds numerically, a parameterization of separable
state is required. Due to the linearity of the trace
and the fact that general separable states are defined as convex mixtures of
pure states, it suffices to consider pure separable states
$\rho_s \in \mathcal{H}_1 \otimes \mathcal{H}_2$ to maximize/minimize
$\tr[\rho_s W]$. Any such state can be generated by a local unitary transformation of
the state $\ketbra{00}{00} \equiv \ketbra{0}{0} \otimes \ketbra{0}{0}$:
\begin{widetext}
\begin{gather}
  \label{paramOfStates}
  \rho_s = \rho_1 \otimes \rho_2 = U_1 \ketbra{0}{0} U_1^\dagger \otimes U_2
  \ketbra{0}{0} U_2^\dagger = U_1 \otimes U_2 \ketbra{00}{00} U_1^\dagger
  \otimes U_2^\dagger \equiv U \ketbra{00}{00} U^\dagger
\end{gather}
\end{widetext}
In Ref.~\cite{spenglerHuberHiesmayr2} an efficient parameterization of unitaries
was proposed that requires only $2(d-1)$ parameters to construct any pure state
$\rho_{1/2}$. Consequently, defining a separable, bipartite state $\rho_s$
requires $4(d-1)$ parameters. Naturally, also mixtures of $K$ pure states can be
generated accordingly by using $4K(d-1)$ parameters to generate $K$ pure states to
be mixed with probabilities $p_1, \dots, p_K$. The optimization over all
separable states to find $\rho_{min/max}$ for a given EW $W$ is carried out for
the parameters of the composite parameterization of unitaries according to
(\ref{paramOfStates}). For $d=3$, $8$ independent parameters need to be
optimized in a bounded region. This is done numerically, using the
implementation of the ``Optim'' \cite{optimJulia} package of the ``LBFGS''
algorithm which uses the gradient and an approximation of the Hessian of
$\tr[W \rho]$. To minimize the risk of identifying only a local maximum/minimum,
the optimization procedures is performed $50$ times with random starting points of
the algorithm and the overall minimum/maximum is taken for $L/U$.

\end{document}